# Localized modes revealed in Random Lasers

**Bhupesh Kumar,**[1] **Ran Homri,**[1] **Priyanka**[1] **Santosh Maurya**[1]**, Melanie Lebental**[2] **and Patrick Sebbah**[1,3,*]

[1]*Department of Physics, The Jack and Pearl Resnick Institute for Advanced Technology, Bar-Ilan University, Ramat-Gan, 5290002 Israel*
[2]*Laboratoire Lumiére matiére et interfaces (LuMin) ENS Paris-Saclay, CNRS, Université Paris-Saclay, CentraleSupelec, 91190, Gif-sur-Yvette, France*
[3]*Institut Langevin, ESPCI ParisTech CNRS UMR7587, 1 rue Jussieu, 75238 Paris Cedex 05, France*
[*]*patrick.sebbah@biu.ac.il*

**Abstract:** In sufficiently strong scattering media, light transport is suppressed and modes are exponentially localized. Anderson-like localized states have long been recognized as potential candidate for high-Q optical modes for low-threshold, cost effective random lasers. Operating in this regime remains however a challenge since Anderson localization is difficult to achieve in optics and nonlinear mode interaction compromise its observation. Here, we exhibit individually each lasing mode of a low-dimension solid-state random laser by applying a non-uniform optical gain. By undoing gain competition and cross-saturation, we demonstrate that all lasing modes are spatially localized. We find that selective excitation reduces significantly the lasing threshold while lasing efficiency is greatly improved. We show further how their spatial location is critical to boost laser power-efficiency. By efficiently suppressing spatial hole burning effect, we can turn on the optimally-outcoupled random lasing modes. Our demonstration opens the road to the exploration of linear and nonlinear mode interactions in the presence of gain, as well as disorder-engineering for laser applications.

## 1. Introduction

Light transport in strongly scattering disordered systems is governed by the nature of the underlying eigenmodes, more specifically, their spatial extension within the scattering medium. For sufficiently strong disorder, the modes of vibration become spatially confined and transport is impeded. Disorder-induced localization is one of the most striking and puzzling manifestation of wave interference, predicted by P.W. Anderson for electrons [1] and later generalized to classical waves [2, 3]. If Anderson localization in three-dimension photonics has never been demonstrated and its very existence is today in question [4], optical localization in lower dimensions is possible for sufficiently large disordered systems. A major challenge however in the study of Anderson-like localized states in low dimensional optical systems resides in the difficulty to excite them independently and observe them individually. Only in the particular situation of extremely strong scattering where modes are well separated and long-lived, have the localized modes been observed individually in optics [5].

A possible route suggested theoretically to overcome this difficulty is to introduce gain and investigate rather random lasing modes [6, 7]. The observation of sharp lasing peaks in semiconductor powders [8] has triggered the interest for random lasers, raising the question of the origin of the laser oscillations and the nature of the lasing modes in these mirrorless lasers [9]. In the regime of Anderson localization, the lasing modes are predicted to build up on the Anderson-localized modes of the passive system [7], although this was never confirmed experimentally. This regime has been proposed to obtain low threshold and stable multimode random lasers [6]. It was further realized that random lasing can also occur in diffusive samples, where modes are spatially extended [10]. Since then, random lasing has been observed and studied in a large variety of systems ranging from $\pi$-conjugated polymers [11] to optofluidic systems [12], from fiber lasers [13–15], bio-polymer film [16] and silkworm silk fibers [17] to planar nanophotonic network [18], to cite a few. But in most cases, scattering was not strong enough to operate in the localized regime. Or at least, localization could not be demonstrated directly. In this context, an unambiguous experimental demonstration of random lasing in the regime of Anderson localization was yet missing. It was first investigated in a stack of microscope cover slides with active dye [19], and later in Er-doped random Bragg grating fiber [20], where a low-threshold laser emission was assumed to occur on long-lived modes overlapping spatially with the gain region, yet without identification of the nature of the modes. Lasing modes localized over only a few camera pixels were directly observed in disordered photonic crystal waveguides embedded with quantum dots [21], while the cavity mode volume was inferred in [22] from the laser input-output characteristic, by relying on a theoretical model of semiconductor laser rate equations [23]. Other indirect methods have been proposed earlier to retrieve modal extension, first in [24] with spectrally-resolved speckle techniques, and in [25–27] by displacing the pump and monitoring the emission spectrum. In contrast, the exponentially-decaying spatial profiles of lasing modes has been directly observed in a jet of dye droplets, using hyper-spectral imaging [28]. However, beside the specificity of the photonic system, the method itself cannot isolate the modes in a regime of strong mode overlap where spatial hole burning effect and mode competition for gain are expected to strongly modify the nature of the modes. A method that truly isolate modes is therefore indispensable to investigate their features. The experimental challenge resides not only in the highly multimode nature of random lasers, but also in the complex association of a non-Hermitian scattering system with a strongly nonlinear active medium. Linear and nonlinear modal interaction, strong gain-induced nonlinearities, including gain saturation and cross-saturation, expected in random lasers makes it even more difficult to extract the actual mode profile from any measurement.

In this article, we investigate random lasing in a strongly scattering active medium and demonstrate disorder-induced Anderson localization of lasing modes. To address this issue, we design an optically-pumped scattering random laser, which allows direct imaging of the field intensity distribution within the laser. We propose to pump non-uniformly the random laser to force it in singlemode operation. We expect the spatially-modulated pump intensity to excite selectively a particular lasing mode, while rejecting all others below threshold. To find the correct pump profile, if it exists, we inspired ourselves from a method recently proposed based on gain shaping and iterative optimization algorithm [29]. This method has been successfully demonstrated experimentally to select modes in a weakly scattering random laser [30], in a micro-disk laser [31] and in asymmetric resonant cavities [32]. It was also proposed to control spectral characteristic [29, 30, 33], laser power efficiency [34], emission directionality [31, 35], laser threshold [36], and modal interactions [37, 38]. Here, we apply this method to select lasing modes individually and image them spatially.

Remarkably, the natural complexity of the active disordered system is disentangled, revealing the localized nature of each lasing modes individually. We test our method by measuring the intensity profile of a large number of laser modes and by computing the probability distribution of the inverse localization length. We demonstrate how hole burning interactions are efficiently suppressed [39], resulting in reduced threshold and large enhancement of slope efficiency. We show how the location of the localized lasing mode is critical to laser power-efficiency by identifying the optimally-outcoupled modes of the random laser [34]. These findings open a new route to a more systematic exploration of optical localized modes in photonics. This includes the investigation of linear and nonlinear modal interactions [38, 40, 41] or the role of nonlinearities on localization [42–44], as well as engineering modal confinement to control mode competition [37, 45] or to explore new states of disorder matter in the presence of gain [46].

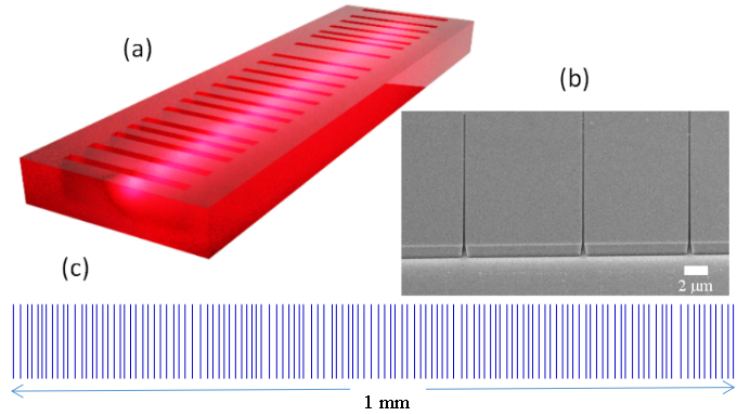

Fig. 1. The solid-state random laser is based on a (DCM) dye-doped 600 nm-thick layer of polymer (PMMA) spin-coated on fused silica. The scattering is provided by 125 parallel grooves (600 nm-deep, 350 nm-wide on average, 50 µm long) randomly distributed along the 1 mm-long sample with an average period of 8 µm. (a) Artist view of the random laser, which is optically pumped from below by a laser strip line. (b)After removing a layer of polymer along the length of the sample, high-resolution scanning electron microscope (SEM) image of the photonic system taken at 52° angle, showing a cross-section of the carved air-grooves. Part of the polymer layer has been removed to show the cross section of the groove (c) Lithographic mask for a particular disorder realization.

## 2. Results

We have designed a solid-state organic random laser (Fig. 1a), which benefits from laser stability, sample longevity and most importantly for our study, relatively strong scattering, a necessary condition to observe the regime of Anderson localization in a finite-size sample. Once polymerized, a 600 nm-thick layer of doped-polymer (DCM-doped PMMA) spin-coated on a fused silica plate, is structured by engraving hundreds of parallel randomly-spaced air-grooves, using electron-beam lithography. Controlled disorder is introduced by varying the groove position around a periodic arrangement. A typical mask and a section of the sample are shown in Fig. 1c and b. SEM image is taken after removing a layer of polymer along the sample length to have a clear view of the carved grooves cross-section. The fabrication process is inspired from the method developed for organic microlasers in [47] and is described in full details in the Supplement 1(section A). When optically-pumped (from the below) at 532 nm, spontaneously emitted light guided within the polymer film is scattered at the grooves interface and amplified as it propagates perpendicular to them, along the sample length. Finally, the degree of scattering and, consequently the localization length, can be adjusted by varying the depth of the grooves. In the present study, the scattering system is composed of 125 identical parallel grooves, uniformly distributed around an 8 µm-pitch periodic lattice with deviation of ± 3 µm (Fig. 1c). Each groove is 600 nm-deep, 50 µm-long, and 350 nm-wide. The total length of the sample is L=1000 µm. Although a similar design is possible in two dimensions, quasi-1D geometry has been preferred here to increase the return-probability and the chances to observe localization within the geometric limit of our sample.

We estimate numerically the localization length in our sample, using a simplified 1D-layered model and the transmission matrix method. The localization length $\xi$ is obtained from the average over a large collection of disordered configurations of the logarithm of the transmission, $\xi = \langle \ln T(L) \rangle$, where $L$ is the sample length (for details see Supplement 1( Section B)). It is found to be $\xi = 65$ μm, well within the geometric limits of our sample. We also performed a more realistic full-2D numerical modeling of the passive system, based on finite-element method (COMSOL-Multiphysics with radio frequency (RF) module), which includes all sample layers and groove geometry.

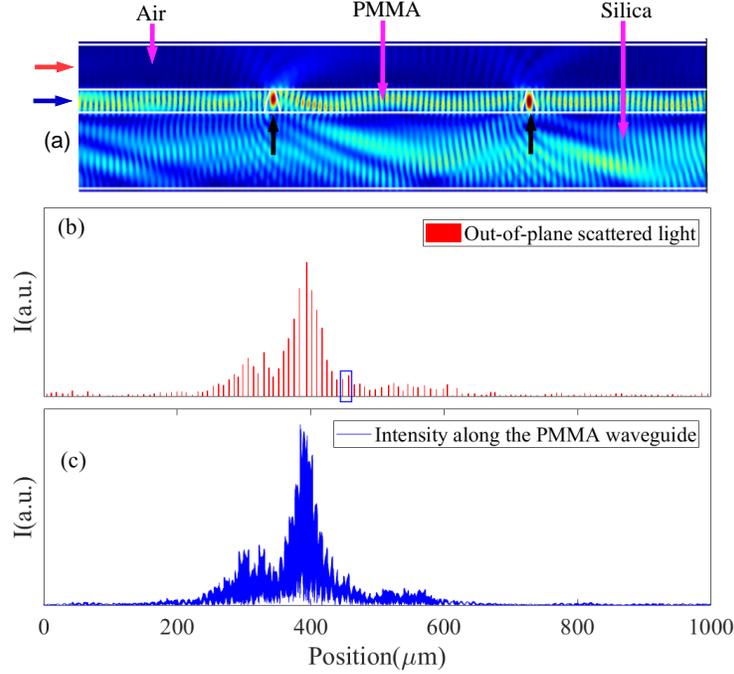

Fig. 2. Full-2D numerical simulation of the field intensity distribution (first mode) of the resonance at $\lambda_0$=601 nm (frequency $f = 4.99\times 10^{14} + i9.7\times 10^{11}$ Hz), calculated for a 1000 μm-long sample with 125 grooves randomly distributed. The sample geometry is reproduced. Thickness of PMMA/DCM layer, air layer and silica substrate are respectively 600 nm, 1200 nm, and 2000 nm. Their respective refractive indices are $n = 1.54$, $n_{air} = 1$, and $n_{Silica} = 1.44$. We use perfectly matched layer (PML) after air and silica substrate. (a) Electric field intensity distribution computed in the vicinity of two slanted grooves, upward black arrow mention position of grooves (identified by the blue box in (b)). (b) Intensity profile of the light scattered by the grooves calculated just above waveguide (red arrow). (c) Intensity profile along the waveguide (blue arrow) for the localized mode shown in (b).

The field-intensity distribution along the waveguide is shown in Fig. 2c. A close-up view reveals the complexity of both in-plane and out-of-plane scattering. The structuration of the sample offers actually an extra bonus: Out-of-plane scattering into air allows imaging of the laser intensity profile close to the sample surface, otherwise confined by the waveguide geometry. Here we can safely assume that this image reproduces the in-plane intensity within the sample. This is confirmed in Figs. 2a and 2b which compare the intensity distribution of a particular localized mode within the waveguide (Fig. 2c) to the intensity scattered by the groove, just above the sample (Fig. 2b). The image of the scattered field-intensity accurately reproduces the intensity profile of the mode. It is therefore possible to retrieve the spatial profile of the mode from the measurement of the out-of-plane scattered light, a useful advantage offered by this system.

A laser strip (500 μm × 50 μm) from a frequency-doubled Nd:YAG laser at 532 nm is used to pump uniformly the random structure. The experimental setup is described in Fig. 3. At laser energy above 18 nJ, multiple scattering provides the necessary optical feedback to reach laser oscillations and sharp discrete peaks appear on a broad spontaneous emission(SE) background ( inset Fig. 4c). As it is most often the case with random lasers, emission is multimode. The out-of-plane laser radiation is imaged under a x10 microscope objective (Fig. 4a), showing out-of-plane scattered light at the position of the air-grooves, in contrast to uniformly-distributed SE detected below threshold (Fig. 4b).

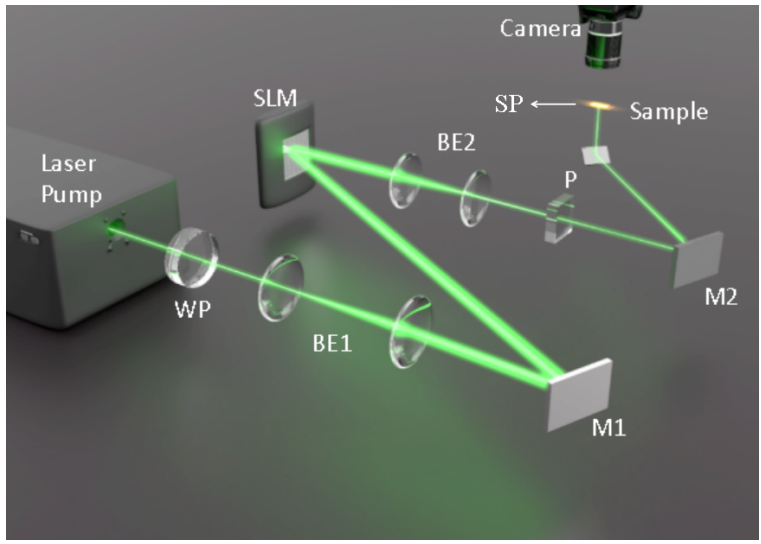

Fig. 3. Schematics of the experimental setup: The 532 nm beam of a picosecond (28 ps) frequency-doubled mode-locked Nd:YAG laser is first expanded on a computer-driven LCOS amplitude spatial light modulator (SLM), which offers the possibility to modulate spatially the pump intensity. The reflected laser pattern is de-magnified and imaged on the back of the sample. A fixed-stage optical microscope (not represented) magnifies (10x) the field-intensity scattered at the surface of the sample onto a sCMOS camera. WP: Half-wave plate; M1 and M2: Mirrors; BE1 and BE2: Beam expanders; P: Polarizer, SP:spectrometer.

By monitoring independently and simultaneously laser intensity and SE, we observe the clamping of spontaneous emission(SE). This feature, which occurs when stimulated emission becomes the preferred transition mechanism, is a fundamental signature of lasing [48, 49]. To the best of our knowledge, it has never been observed for random lasers. It is shown in Fig. 4b and c, where the growth of SE is seen to be drastically curtailed as the threshold is reached and laser oscillations take over.

The field intensity distribution is recovered after subtracting the spontaneous-emission(SE) contribution. This procedure is detailed in the Supplement 1 (Section C). A typical intensity profile of laser emission when uniformly-pumped is shown in Fig. 5h, corresponding to the multimode emission spectrum shown in Fig. 5a. Several modes contribute to this profile, preventing any investigation of their spatial distribution individually. To observe them separately, we propose to operate the random laser in single mode at the lasing wavelength we want to investigate. This is achieved by tailoring the intensity profile of the pump in order to excite selectively this particular mode. After reflection on a computer-driven spatial light modulator (SLM), the pump beam is modulated and projected on the back-side of the sample (see Fig. 3 and Supplement 1 (Section D). Because the pump profile that selects a particular mode is not known a *priori*, an iterative method based on an optimization algorithm is used following [30] (see Supplement 1( Section E, F)). We first aim at selecting a mode which is not the first mode to lase under uniform pumping at $\lambda$ = 599.80 nm. After 200 iterations, a non-uniform pump profile is found (top inset of Fig. 5i), which optimally selects the targeted mode lasing wavelength and rejects all others below threshold, as demonstrated in the Fig. 5b. Remarkably, pure single mode operation is achieved, in contrast to [30] where mode rejection was not total. The out-of plane scattered light is imaged from the top of the sample. The resulting intensity profile is shown in Fig. 5i. The lasing mode is found to be spatially confined away from the sample limits. This indicates that random lasing indeed occurs in the localized regime. The same operation is reproduced to select other lasing modes from the multimode emission spectrum of Fig. 5a. For each mode, a new pump profile is systematically found which selects this particular mode. Imaging the selected modes reveals that all lasing modes are spatially localized. Fig. 5 (b-g) shows emission spectra of six individually selected lasing mode and corresponding lasing mode intensity distribution, together with their optimized pump profile shown in Fig. 5 (i-n). Successive optimization operations usually lead to different pump profile. Remarkably however, the spatial intensity profile obtained after optimization is found to be unique and independent of the optimized pump profile ( for details see Supplement 1 (Section G)).

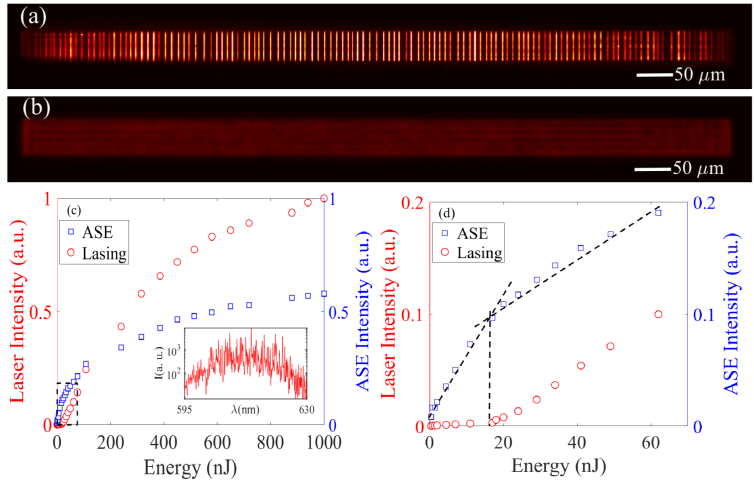

Fig. 4. Optical microscope image of field-intensity distribution near the sample surface when pumped (a) above threshold and (b) below threshold. (c) Spontaneous emission(SE) (blue open square) and scattered laser intensity (red open circle) vs. pump energy. inset: emission spectrum with sharp discrete peaks on a broad SE background plotted in semi-log scale. (d) Zoom-in of (c). Clamping of the SE is observed at threshold, for a pump energy of 18 nJ. Note that the laser intensity measured here from out-of-plane scattering is only a small fraction of the actual laser emission, which explains that it is comparable to SE.

Lasing modes are found to be exponentially localized in different regions of the sample and occupy a significant fraction of the sample length. Despite strong spatial overlap, it is possible to measure their spatial extension. The relatively easy access to the spatial distribution of individual localized modes allows us to explore the probability distribution of the localization length of the eigenstates. Its inverse, which is called the Lyapunov exponent, is expected to follow a normal distribution in a sufficiently long 1D random system with weak disorder [50]. Deviation from normal distribution, when far tail becomes significant, can be induced by e.g. the different nature of the states [51], disorder correlation [52] or absorption [53]. Such a deviation has been at the center of a strong debate, as it is directly related to the breakdown of the single-parameter scaling theory, which is one of the foundations of localization theory [54]. Here, we measured the mode extension for 75 lasing modes individually-selected in 10 samples with different disordered configurations. All samples were carefully manufactured under identical conditions. We define the mode's length as the distance between the extreme scattering grooves from which intensity of light scattered is above noise level. The inverse mode-length [50] probability distribution is plotted in Fig.6, together with a normal fit. Tail to the distribution indicates departure from normal distribution. The half of the mean-value of the distribution gives the average localization length, $\xi$ =100 $\mu$m. Inset in Fig.6 shows the intensity profile of a localized mode in log-scale. Exponential decay is not always as sharply defined as in this case. We use therefore the spatial mode extension to extension to estimate localization length.

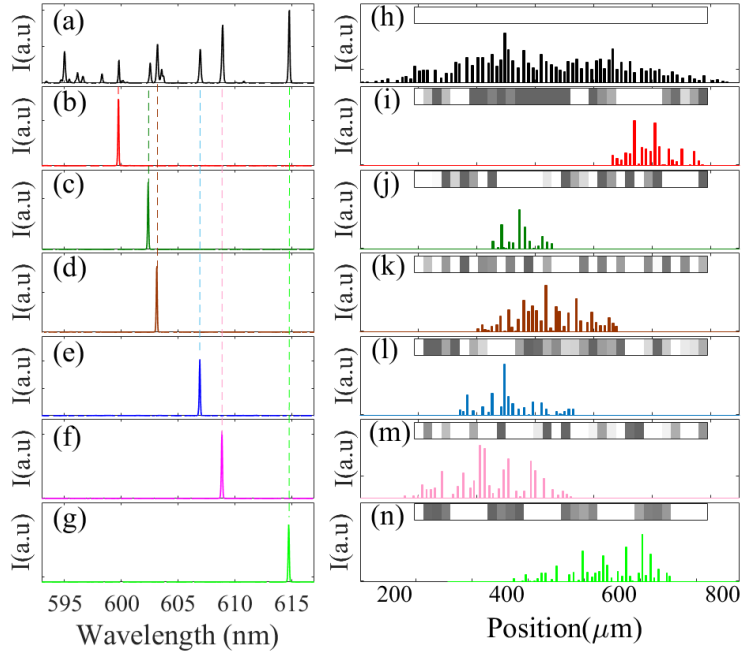

Fig. 5. Selective pumping of localized lasing modes. (a) Random laser emission spectrum obtained in the case of uniform pumping (pump energy is 71.2 nJ). (b-g) Emission spectrum of selected modes after optimization at target wavelengths: $\lambda$ =599.78 nm, 602.44 nm, 603.17 nm, 606.95 nm, 608.87 nm and 614.75 nm respectively. (h) Spatial field-intensity distribution corresponding to emission spectrum in (a), measured at the groove positions. Each bar integrates light-intensity scattered by one groove. The pump intensity profile is shown on top. (i-n) Spatial distribution of laser intensity within the random structure corresponding to each individually selected mode shown in left panel. The corresponding optimized pump profile is reproduced in grayscale on top of each figure (white: maximum intensity; Black: no pumping). The spectral mode rejection w.r.t. maximum noise level in the spectra obtain with this method is 56.18 dB, 64.49 dB, 60.56 dB, 58.72 dB, 66.41 dB and 69.87 dB, respectively.

We further point out here that the optimization algorithm does not trivially converge to a narrow pump profile near the center of the mode, which would have been sufficient to select spatially-isolated and strongly confined modes, [7, 29]. Here, the distribution of the pump intensity extends far beyond the center of the mode. The reason for this non-trivial optimal pump profile is that modes overlap significantly in our system and therefore compete for gain. Optimal selection of a single lasing mode necessitates therefore to disentangle interacting modes. This is achieved by suppressing hole burning effect and turning off cross-saturation. The ability of pump shaping to isolate each mode individually is remarkably demonstrated in Fig. 7, which compares the laser characteristic of the mode @599.78 nm for uniform and optimized pump profile. When the system is pumped uniformly (multimode regime), the targeted-mode emission @599.78 nm saturates rapidly, impeded by mode competition and cross-saturation between different lasing modes. Once optimized, the laser intensity becomes linear with pump energy, confirming that no other mode is lasing simultaneously. The selecting profile is found to be efficient in preventing spatial hole burning, below and far above the pump energy at which the optimization was performed. Strikingly, the threshold of the lasing mode after optimization is reduced by a factor 1.5, while the slope efficiency of its laser characteristic is increased by a factor six, when compared to the same mode lasing in multimode regime under uniform pumping. This means that the energy couples out of the random laser more efficiently when the mode is optimized. The selective method employed here unleashes the selected mode from competing for gain with other modes. This is an important condition to investigate the nature of localized states, which other methods, including hyperspectral imaging, may not allow when mode overlap is important. The performance of the laser has therefore been improved by "focusing" the pump where it is needed and consequently reducing gain surface, a rather counter intuitive effect.

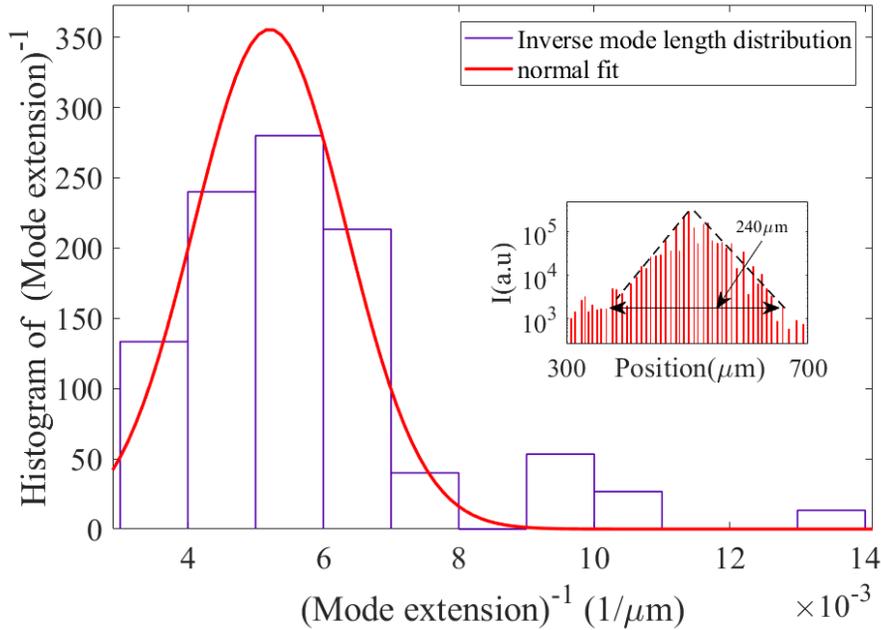

Fig. 6. Histogram of the inverse mode spatial extension probability distribution calculated for 75 different lasing modes, selected individually. solid red line is normal fit. Inset shows spatial field distribution of a localized mode in log scale.

This significant improvement in laser energy-efficiency can be pushed even a step further, by selecting the optimally-outcoupled lasing mode of the random laser. The optimally outcoupled lasing mode of an optical resonator is the mode that provides the largest slope efficiency. It was argued in [34] that such modes are suppressed by spatial hole burning interactions in a uniformly-pumped laser. Shaping the pump profile was therefore proposed [34, 39] to control spatial hole burning, increase laser stability and laser power-efficiency, and to allow lasing in the most efficient mode. This was tested numerically in a microdisk laser, where the optimally-coupled mode can be calculated and the pump profile is trivial. In a random laser however, both modal identification and pump selection turn out to be much more challenging. Here, we show how our approach, which couples mode selection and mode imaging, allows to identify the optimally outcoupled modes of our random laser. We measure the threshold and the slope efficiency of 15 individual lasing modes of a given sample, after they have been selectively pumped. Threshold and slope efficiency are then correlated with the position of each mode along the sample, defined here as the position of its barycenter. The laser threshold (red plot in Fig. 8) is seen to grow for modes near the sample boundaries, as leakage out from the scattering medium increases the loss of these modes. Actually, the measure of the threshold provides a direct evaluation of the sensitivity of the localized modes to the boundary of the system [55]. The slope efficiency presented in blue plot in (Fig. 8) shows a remarkable feature: It exhibits two maxima at positions 200 $\mu$m and 700 $\mu$m, almost symmetric around the center of the sample. Maximum slope efficiency defines the optimally-outcoupled lasing modes for this particular disordered system. It is worth noting that this happens for lasing modes neither near the center of the sample, nor near the sample edges. This shows how critical is the spatial position of the optimally-outcoupled modes. It is a delicate trade-off between low-loss modes, which build up energy near the center, and efficiently output-coupled modes close to the sample edges. This demonstrates how mode selection by tailoring the pump profile can turn on the most efficient lasing modes of a particular random laser. Our findings extend the theoretical predictions in [34] to the most challenging case of random lasers.

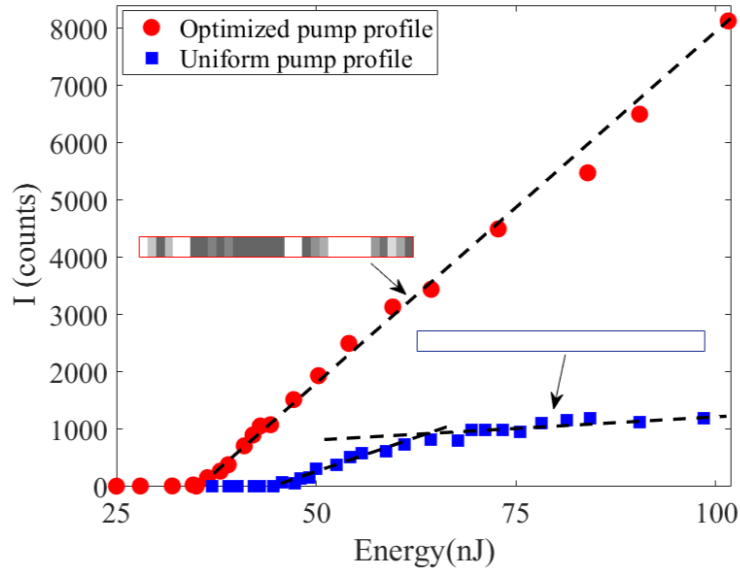

Fig. 7. Laser characteristic (Intensity vs pump energy) for the lasing mode @599.78 nm: (blue full square) In multimode operation (uniform pump profile); (red full circle) In singlemode operation (optimized pump profile). In multimode regime, the threshold of the mode @599.78 nm is 47.3 nJ and its slope efficiency is 20 counts/nJ. In single mode regime, its threshold reduces to 35 nJ, and its slope efficiency increases to 123 counts/nJ.

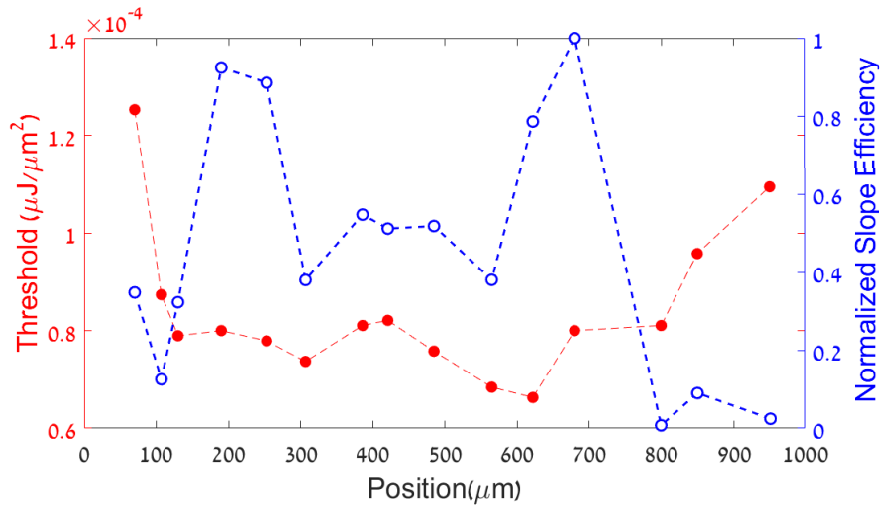

Fig. 8. Lasing threshold (red full circles) and normalized slope efficiency (blue open circles) vs. mode position along the sample of individually-excited lasing modes. From left to right: $\lambda$= 622.12 nm, 609.02 nm, 605.05 nm, 606.91 nm, 595.48 nm, 604.73 nm, 606.49 nm, 619.04 nm, 614.48 nm, 612.60 nm, 617.15 nm, 606.35 nm, 598.66 nm, 602.74 nm. The position of the mode is defined as the position of its barycenter.

## 3. Conclusion

In conclusion, we have explored random lasing in the strongly scattering regime, and demonstrated disorder-induced localization, by combining selective pump shaping and imaging of individual lasing modes in a newly-designed random laser. By coupling spatial imaging to the selective pumping method, we confirmed the spatial confinement of all lasing modes and we identified the optimally outcoupled modes of the random laser. Selective excitation by non-uniform pumping proves itself remarkably efficient to suppress spatial hole burning, modal interaction and cross-saturation effect in an-otherwise strongly non-Hermitian system. We claim that this method works actually as an effective eigensolver to retrieve the localized modes individually and to identify the lasing modes which boost laser efficiency, which is out of reach to other methods found in literature. Our findings can be extended to 2D-systems, where localized laser modes are more challenging to achieve. It opens a unique route to investigate Anderson localization, to explore the role of nonlinearities on localization and test experimentally theoretical predictions [42]. It can be applied also to the design of highly efficient and stable random microlasers, where the random and non-Hermitian nature of these lasers offers unprecedented degrees of freedom.

**Acknowledgments.** We thank Profs. V. Freilikher and A. Z. Genack for useful discussions. We are grateful to Dr. Yossi Abulafia for his help in the fabrication process and to Bar-Ilan Institute of Nanotechnology & Advanced Materials for providing with fabrication facilities.

**Data availability.** Data underlying the results presented in this paper are not publicly available at this time but may be obtained from the authors upon reasonable request.

**Disclosures.** The authors declare no conflicts of interest.

See Supplement 1 for supporting content.

# Localized modes revealed in Random Lasers: Supplementary material

Bhupesh Kumar,[1] Ran Homri,[1], Priyanka[1], Santosh Maurya [1], Melanie Lebental[2] and Patrick Sebbah[1,3,*]

[1]*Department of Physics, The Jack and Pearl Resnick Institute for Advanced Technology, Bar-Ilan University, Ramat-Gan, 5290002 Israel*

[2]*Laboratoire Lumiére matiére et interfaces (LuMin) ENS Paris-Saclay, CNRS, Université Paris-Saclay, CentraleSupelec, 91190, Gif-sur-Yvette, France*

[3]*Institut Langevin, ESPCI ParisTech CNRS UMR7587, 1 rue Jussieu, 75238 Paris Cedex 05, France*
*\**patrick.sebbah@biu.ac.il*

## This document provide supplementary information to " Localized modes revealed in Random Lasers "

### A. Device Fabrication.

To fabricate our samples, we use PMMA (polymethyl methacrylate, microchem,USA) with a molecular weight of 495000 g/mol at a concentration of 6% weight in anisole. It is doped with 5% weight of DCM(Exciton) laser dye (4-dicyanomethylene-2-methyl-6-(4-dimethylaminostyryl-4H-pyran). DCM is preferred because it has a fluorescence spectrum centered around 600 nm with a good quantum yield and a large stoke shift (100 nm), which prevents reabsorption of the emitted light. The mixture is spin-coated at 1000 rpm for 60 sec and post baked at 120 ºC for 2 hours in oven, to obtain a 600 nm-thick layer a 1 mm-thick fused-silica plate (Edmund optics). The refractive index of the silica plate is 1.45, compare to 1.54 for PMMA-DCM layer. Disordered structures are carved using electron-beam lithography (CRESTEC/CABL-9000C). Electron-beam exposure on bare PMMA-DCM layer generates significant charging effects, as expected in case of insulating substrate. To avoid that, we add a 20 nm layer of conductive polymer (E-Spacer) to eliminate the charging effect. After e-beam writing, the conductive polymer is removed with de-ionized water by immersing it for 40s. The resulting grooves are 350 nm-wide, 50 $\mu$m-long. They are separated on-average by 8 $\mu$m of doped-PMMA (see Fig.1(c)). This averaged pitch has been chosen to bring the localized modes in the spectral line of the dye emission. The localization length depends on the depth and spatial distribution of the grooves. The width and depth of air grooves can be varied by changing current and exposure time of electron beam. The fabrication process is based on the method developed in [1].

### B. Computing the localization length

To calculate the localization length, we consider a simple structure similar to our real sample. Disorder with standard deviation($\eta$) of 3 $\mu$m is added to the mean position of each groove of a periodic structure of pitch(P) 8 $\mu$m using the formula $d_i = P + (2*\zeta)*\eta$, where $\zeta$ is a uniformly distributed random numbers between [1,-1] generated using the random number generator *rand* of MATLAB. Air groove width is 350 μm. Refractive index of dielectric layer is $n_2$ =1.54 while it is $n_1$ =1 for air. Refractive index outside the disorder structure is kept at $n = 1$. Transfer matrix method developed e.g.in [3] is used to calculate $\langle \ln T(L) \rangle$ for several thousands of disorder configurations. where T is transmission coefficient and L is length of the system. Localization length is given by:

$$\xi^{-1}= -d \langle \ln T(L) \rangle/dL$$

and averaged over wavelength range of interest.

### C. Baseline Correction

We find that, below threshold, when spontaneous emission dominates, out-of-plane scattering from grooves is negligible. As pump energy exceeds laser threshold, we observe very sharp out-of-plane light scattering from the grooves. Fig.S1(a) shows transversely integrated profile of uniformly-pumped sample. The intensity profile, integrated over the width of the sample, consists of both SE and lasing intensity. Spline fitting is used to remove SE part from the total integrated intensity profile. Fig.S1(b) shows the SE alone, while Fig. S1(c) shows out-of-plane scattering due to random lasing. Each peak corresponds to scattering form one air groove. Fig.S1(d) shows the intensity distribution in bar plots, each bar representing peak intensity, integrated over 3 pixels.

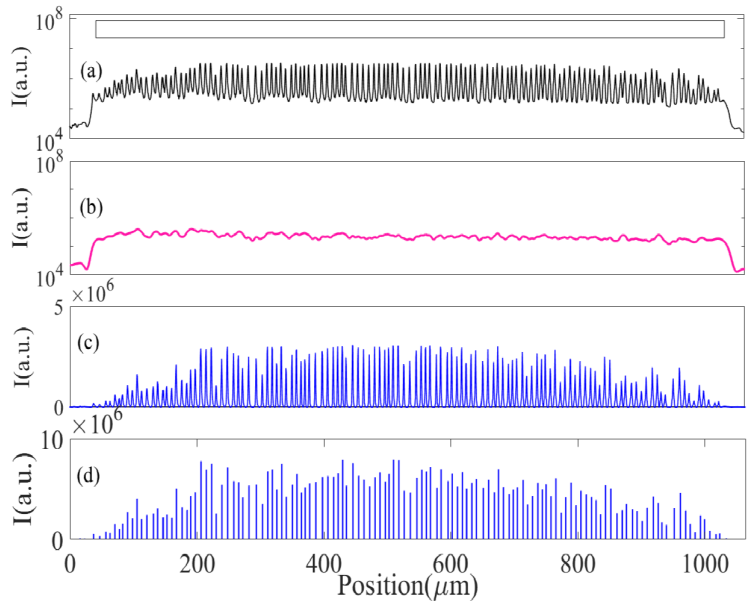

Fig. S1. (a) Semi-log plot of transverse intensity profile integrated over the width (50$\mu$m) of the 1D disorder sample far above threshold uniform pumping. Top inset shows uniform pump profile. (b) Semi-log plot of SE component extracted from (a) using spline fitting. (c) Sharp intensity peaks correspond to the lasing light scattered from the air grooves. (d) Corresponding bar-plot representation:each bar representing peak intensity, integrated over 3 pixels.

### D. Experimental Set-up.

A schematics of the experimental setup is shown in Fig.3. The pump beam from a frequency-doubled Nd:YAG laser (EXPLA PL2230: 532 nm, 20 ps, maximum output energy 28 mJ, repetition rate 10 Hz) is expanded 5x to be spatially modulated by a 1952×1088-pixels reflective spatial light modulator (SLM) (Holoeye HES 6001, pixel 8.0 $\mu$m). The SLM is placed in the object plane of a telescope with 5x reduction and is imaged on the back of the sample. This setup provides a 500 $\mu$m-long and 50 $\mu$m-wide laser strip line with nearly diffraction-free modulation. The random laser sample is precisely aligned with the pump laser strip line under a fixed-stage

Zeiss (AxioExaminer A1) microscope and imaged using an Andor Zyla sCMOS camera (22 mm diagonal view, 6.5 $\mu$m pixel size) attached to the microscope. In-plane laser emission exiting from one end of the device is focused by a microscope objective (20x) into the tip of a multimode fiber coupled to a high-resolution imaging spectrometer (Horiba iHR550), equipped with a 2400 mm$^{-1}$ grating and a back-illuminated deep-depleted thermoelectrically-cooled CCD camera (Synapse CCD with sampling rate 1 MHz, 1024 X 256 pixels, 26 $\mu$m pixel pitch). We confirm that spectra measured at the end of the sample or from its top via out-of-plane scattered light are similar (same peaks position but different heights, depending on the position of observation). The entrance slit of the spectrometer is 50 $\mu$m. The resulting spectral resolution is 20 pm. The integration time is 1sec.

### E. Optimization algorithm [2]

We use Nelder mead simplex method executed in the *fminsearch* function of Matlab. We modified this function by setting the initial step (usual $\delta$ parameter in *fminsearch*) to 1.0 to explore large regions of the 32 dimensional space. The number of pixel 32 has been chosen for the best compromise between sensitivity and computation time.

### F. Optimization method [2]

The Matlab-generated image send to the SLM is made up of 32 intensity blocks. Each intensity block is greyscale-coded on 256 levels and projected onto the sample. We choose 32 column-vectors $V_i$ of the 32 × 32 binary Hadamard matrix as the initial vertex to start the optimization method.

The pump profile P(x) is therefore written as:

$$P(x) = \frac{1}{255} \sum_{i=1}^{32} \beta_i V_i$$

where $\beta_i$ takes discrete values in the range [0,255]. Each vector $\beta_i$ corresponds to a particular pump profile associated with a particular emission spectrum $I(\lambda)$. To achieve single mode operation at a desired wavelength $\lambda_0$ we need to find the vector, which maximizes the ratio $R(\lambda_0)=I(\lambda_0)/I(\lambda_1)$, where $\lambda_1$ corresponds to the wavelength of the lasing mode with highest intensity, apart from the mode at $\lambda_0$. Here, $I(\lambda)$ represents the intensity at $\lambda$ after the spectrum baseline has been subtracted. Experimentally, for a given pump profile 10 spectrum are averaged over 1 sec. Typical optimization last 20 minutes.

### G. Robustness of selected localized mode w.r.t. optimization process

We found that the emission wavelength and spatial intensity distribution of any localized mode remain unchanged for different optimized pump profiles obtained after running different optimization process. This is demonstrated here for a particular lasing mode, and confirms for all modes of the system. We run three independent optimization routines to select the mode lasing at $\lambda$ =605.50 nm. Fig. S2 shows the emission spectrum, mode intensity distribution after optimization and the corresponding pump intensity profile. Both the spectral peak position and the spatial intensity distribution of the mode are found to remain unchanged, which confirms that the native distribution of the lasing mode has been identified.

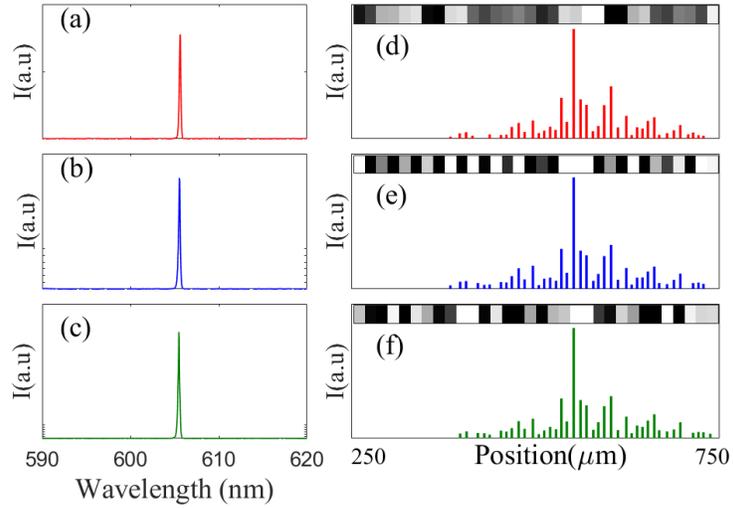

Fig. S2. (a-c) Emission spectra of individually selected mode at $\lambda$= 605.5 nm obtained after three independent optimization routines. (d-f) Spatial intensity profile of the selected mode. The corresponding optimized pump profile is reproduced in greyscale on top of each figure (white: maximum pump intensity; Black: no pumping).